\documentclass[%
aps,
prl,
reprint, %Two column journal format
superscriptaddress,
amsmath,
amssymb,
]{revtex4-2}
\usepackage[breaklinks=true,colorlinks=true,linkcolor=blue]{hyperref}
\usepackage[normalem]{ulem} %for \sout{} (strikeout)
\usepackage[utf8]{inputenc}
\usepackage{amsmath}
\usepackage{graphicx}% Include figure files
\usepackage{dcolumn}% Align table columns on decimal point
\usepackage{bm}% bold math
\usepackage{xcolor}% Text highlighting
\usepackage{booktabs}
\hypersetup{colorlinks=true,            allcolors=blue,            bookmarksopen=false,            bookmarksnumbered}            

\begin{document}

\title{First Determination of the $^{27}$Al Neutron Distribution Radius\\ from a Parity-Violating Electron Scattering Measurement}

%Qweak collaboration
\author{D.~Androi\'c}
\affiliation{University of Zagreb, Zagreb, HR 10002, Croatia }
\author{D.S.~Armstrong}
\email[corresponding author: ]{armd@physics.wm.edu}
\affiliation {William \& Mary, Williamsburg, Virginia 23185, USA}
\author{K.~Bartlett}
\affiliation {William \& Mary, Williamsburg, Virginia 23185, USA}
\author{R.S.~Beminiwattha}
\affiliation{Ohio University, Athens, Ohio 45701, USA}
\author{J.~Benesch}
\affiliation{Thomas Jefferson National Accelerator Facility, Newport News, Virginia 23606, USA}
\author{F.~Benmokhtar}
\affiliation{Christopher Newport University, Newport News, Virginia 23606, USA}
\author{J.~Birchall}
\affiliation{University of Manitoba, Winnipeg, Manitoba R3T2N2, Canada}
\author{R.D.~Carlini}
\affiliation{Thomas Jefferson National Accelerator Facility, Newport News, Virginia 23606, USA}
\affiliation {William \& Mary, Williamsburg, Virginia 23185, USA}
\author{J.C.~Cornejo}
\affiliation {William \& Mary, Williamsburg, Virginia 23185, USA}
\author{S.~Covrig Dusa}
\affiliation{Thomas Jefferson National Accelerator Facility, Newport News, Virginia 23606, USA}
\author{M.M.~Dalton}
\affiliation{University of Virginia,  Charlottesville, Virginia 22903, USA}
\affiliation{Thomas Jefferson National Accelerator Facility, Newport News, Virginia 23606, USA}
\author{C.A.~Davis}
\affiliation{TRIUMF, Vancouver, British Columbia V6T2A3, Canada}
\author{W.~Deconinck}
\affiliation {William \& Mary, Williamsburg, Virginia 23185, USA}
\author{J.F.~Dowd}
\affiliation {William \& Mary, Williamsburg, Virginia 23185, USA}
\author{J.A.~Dunne}
\affiliation{Mississippi State University,  Mississippi State, Mississippi 39762, USA}
\author{D.~Dutta}
\affiliation{Mississippi State University,  Mississippi State, Mississippi 39762, USA}
\author{W.S.~Duvall}
\affiliation{Virginia Polytechnic Institute \& State University, Blacksburg, Virginia 24061, USA}
\author{M.~Elaasar}
\affiliation{Southern University at New Orleans, New Orleans, Louisiana 70126, USA}
\author{W.R.~Falk}
\thanks{Deceased}
\affiliation{University of Manitoba, Winnipeg, Manitoba R3T2N2, Canada}
\author{J.M.~Finn}
\thanks{Deceased}
\affiliation {William \& Mary, Williamsburg, Virginia 23185, USA}
\author{T.~Forest}
\affiliation{Idaho State University, Pocatello, Idaho 83209, USA}
\affiliation{Louisiana Tech University, Ruston, Louisiana 71272, USA}
\author{C.~Gal}
\affiliation{University of Virginia,  Charlottesville, Virginia 22903, USA}
\author{D.~Gaskell}
\affiliation{Thomas Jefferson National Accelerator Facility, Newport News, Virginia 23606, USA}
\author{M.T.W.~Gericke}
\affiliation{University of Manitoba, Winnipeg, Manitoba R3T2N2, Canada}
\author{V.M.~Gray}
\affiliation {William \& Mary, Williamsburg, Virginia 23185, USA}
\author{K.~Grimm}
\affiliation{Louisiana Tech University, Ruston, Louisiana 71272, USA}
\affiliation {William \& Mary, Williamsburg, Virginia 23185, USA}
\author{F.~Guo}
\affiliation{Massachusetts Institute of Technology,  Cambridge, Massachusetts 02139, USA}
\author{J.R.~Hoskins}
\affiliation {William \& Mary, Williamsburg, Virginia 23185, USA}
\author{D.C.~Jones}
\affiliation{University of Virginia,  Charlottesville, Virginia 22903, USA}
\author{M.K.~Jones}
\affiliation{Thomas Jefferson National Accelerator Facility, Newport News, Virginia 23606, USA}
%\author{R.~Jones}
%\affiliation{University of Connecticut,  Storrs-Mansfield, Connecticut  06269, USA}
\author{M.~Kargiantoulakis}
\affiliation{University of Virginia,  Charlottesville, Virginia 22903, USA}
\author{P.M.~King}
\affiliation{Ohio University, Athens, Ohio 45701, USA}
\author{E.~Korkmaz}
\affiliation{University of Northern British Columbia, Prince George, British Columbia V2N4Z9, Canada}
\author{S.~Kowalski}
\affiliation{Massachusetts Institute of Technology,  Cambridge, Massachusetts 02139, USA}
\author{J.~Leacock}
\affiliation{Virginia Polytechnic Institute \& State University, Blacksburg, Virginia 24061, USA}
\author{J.~Leckey}
\affiliation {William \& Mary, Williamsburg, Virginia 23185, USA}
\author{A.R.~Lee}
\affiliation{Virginia Polytechnic Institute \& State University, Blacksburg, Virginia 24061, USA}
\author{J.H.~Lee}
\affiliation {William \& Mary, Williamsburg, Virginia 23185, USA}
\affiliation{Ohio University, Athens, Ohio 45701, USA}
\author{L.~Lee}
\affiliation{TRIUMF, Vancouver, British Columbia V6T2A3, Canada}
\affiliation{University of Manitoba, Winnipeg, Manitoba R3T2N2, Canada}
\author{S.~MacEwan}
\affiliation{University of Manitoba, Winnipeg, Manitoba R3T2N2, Canada}
\author{D.~Mack}
\affiliation{Thomas Jefferson National Accelerator Facility, Newport News, Virginia 23606, USA}
\author{J.A.~Magee}
\affiliation {William \& Mary, Williamsburg, Virginia 23185, USA}
\author{R.~Mahurin}
\affiliation{University of Manitoba, Winnipeg, Manitoba R3T2N2, Canada}
\author{J.~Mammei}
\affiliation{Virginia Polytechnic Institute \& State University, Blacksburg, Virginia 24061, USA}
\affiliation{University of Manitoba, Winnipeg, Manitoba R3T2N2, Canada}
\author{J.W.~Martin}
\affiliation{University of Winnipeg, Winnipeg, Manitoba R3B2E9, Canada}
\author{M.J.~McHugh}
\affiliation{George Washington University, Washington, DC 20052, USA}
\author{D.~Meekins}
\affiliation{Thomas Jefferson National Accelerator Facility, Newport News, Virginia 23606, USA}
%\author{J.~Mei}
%\affiliation{Thomas Jefferson National Accelerator Facility, Newport News, Virginia 23606, USA}
\author{K.E.~Mesick}
\affiliation{George Washington University, Washington, DC 20052, USA}
\affiliation{Rutgers, The State University of New Jersey, Piscataway, New Jersey 08854, USA}
\author{R.~Michaels}
\affiliation{Thomas Jefferson National Accelerator Facility, Newport News, Virginia 23606, USA}
\author{A.~Micherdzinska}
\affiliation{George Washington University, Washington, DC 20052, USA}
\author{A.~Mkrtchyan}
\affiliation{A.~I.~Alikhanyan National Science Laboratory (Yerevan Physics Institute), Yerevan 0036, Armenia}
\author{H.~Mkrtchyan}
\affiliation{A.~I.~Alikhanyan National Science Laboratory (Yerevan Physics Institute), Yerevan 0036, Armenia}
\author{A.~Narayan}
\affiliation{Mississippi State University,  Mississippi State, Mississippi 39762, USA}
\author{L.Z.~Ndukum}
\affiliation{Mississippi State University,  Mississippi State, Mississippi 39762, USA}
\author{V.~Nelyubin}
\affiliation{University of Virginia,  Charlottesville, Virginia 22903, USA}
\author{Nuruzzaman}
\affiliation{Hampton University, Hampton, Virginia 23668, USA}
\affiliation{Mississippi State University,  Mississippi State, Mississippi 39762, USA}
\author{W.T.H van Oers}
\affiliation{TRIUMF, Vancouver, British Columbia V6T2A3, Canada}
\affiliation{University of Manitoba, Winnipeg, Manitoba R3T2N2, Canada}
\author{V.F. Owen}
\affiliation {William \& Mary, Williamsburg, Virginia 23185, USA}
\author{S.A.~Page}
\affiliation{University of Manitoba, Winnipeg, Manitoba R3T2N2, Canada}
\author{J.~Pan}
\affiliation{University of Manitoba, Winnipeg, Manitoba R3T2N2, Canada}
\author{K.D.~Paschke}
\affiliation{University of Virginia,  Charlottesville, Virginia 22903, USA}
\author{S.K.~Phillips}
\affiliation{University of New Hampshire, Durham, New Hampshire 03824, USA}
\author{M.L.~Pitt}
\affiliation{Virginia Polytechnic Institute \& State University, Blacksburg, Virginia 24061, USA}
\author{R.W. Radloff}
\affiliation{Ohio University, Athens, Ohio 45701, USA}
\author{J.F.~Rajotte}
\affiliation{Massachusetts Institute of Technology,  Cambridge, Massachusetts 02139, USA}
\author{W.D.~Ramsay}
\affiliation{TRIUMF, Vancouver, British Columbia V6T2A3, Canada}
\affiliation{University of Manitoba, Winnipeg, Manitoba R3T2N2, Canada}
\author{J.~Roche}
\affiliation{Ohio University, Athens, Ohio 45701, USA}
\author{B.~Sawatzky}
\affiliation{Thomas Jefferson National Accelerator Facility, Newport News, Virginia 23606, USA}
\author{T.~Seva}
\affiliation{University of Zagreb, Zagreb, HR 10002, Croatia }
\author{M.H.~Shabestari}
\affiliation{Mississippi State University,  Mississippi State, Mississippi 39762, USA}
\author{R.~Silwal}
\affiliation{University of Virginia,  Charlottesville, VA 22903, USA}
\author{N.~Simicevic}
\affiliation{Louisiana Tech University, Ruston, Louisiana 71272, USA}
\author{G.R.~Smith}
\email[corresponding author: ]{smithg@jlab.org}
\affiliation{Thomas Jefferson National Accelerator Facility, Newport News, Virginia 23606, USA}
\author{P.~Solvignon}
\thanks{Deceased}
\affiliation{Thomas Jefferson National Accelerator Facility, Newport News, Virginia 23606, USA}
\author{D.T.~Spayde}
\affiliation{Hendrix College, Conway, Arkansas 72032, USA}
\author{A.~Subedi}
\affiliation{Mississippi State University,  Mississippi State, Mississippi 39762, USA}
%\author{R.~Subedi}
%\affiliation{George Washington University, Washington, DC 20052, USA}
\author{R.~Suleiman}
\affiliation{Thomas Jefferson National Accelerator Facility, Newport News, Virginia 23606, USA}
\author{V.~Tadevosyan}
\affiliation{A.~I.~Alikhanyan National Science Laboratory (Yerevan Physics Institute), Yerevan 0036, Armenia}
\author{W.A.~Tobias}
\affiliation{University of Virginia,  Charlottesville, Virginia 22903, USA}
\author{V.~Tvaskis}
\affiliation{University of Winnipeg, Winnipeg, Manitoba R3B2E9, Canada}
\affiliation{University of Manitoba, Winnipeg, Manitoba R3T2N2, Canada}
\author{B.~Waidyawansa}
\affiliation{Ohio University, Athens, Ohio 45701, USA}
\author{P.~Wang}
\affiliation{University of Manitoba, Winnipeg, Manitoba R3T2N2, Canada}
\author{S.P.~Wells}
\affiliation{Louisiana Tech University, Ruston, Louisiana 71272, USA}
\author{S.A.~Wood}
\affiliation{Thomas Jefferson National Accelerator Facility, Newport News, Virginia 23606, USA}
\author{S.~Yang}
\affiliation {William \& Mary, Williamsburg, Virginia 23185, USA}
\author{P.~Zang}
\affiliation{Syracuse University, Syracuse, New York 13244, USA}
\author{S.~Zhamkochyan}
\affiliation{A.~I.~Alikhanyan National Science Laboratory (Yerevan Physics Institute), Yerevan 0036, Armenia}

\collaboration{Q$_{\text{weak}}$ Collaboration}%\noaffiliation

\author{M.E.~Christy}
\affiliation{Hampton University, Hampton, Virginia 23668, USA}
\author{C.J~Horowitz}
\affiliation{Indiana University, Bloomington, Indiana 47405, USA}
\author{F.J.~Fattoyev}
\thanks{Now at Manhattan College, Riverdale, New York 10471, USA}
\affiliation{Indiana University, Bloomington, Indiana 47405, USA}
\author{Z.~Lin}
\thanks{Now at University of Tennessee, Knoxville, Tennessee 37796, USA }
\affiliation{Indiana University, Bloomington, Indiana 47405, USA}

%\date{Draft 3.0: \today}

\begin{abstract}
We report the first measurement
of the parity-violating elastic electron scattering asymmetry on $^{27}$Al.  
The $^{27}$Al elastic asymmetry is $A_{\rm PV} = 2.16\pm0.11\; \text{(stat)}\pm0.16\; \text{(syst) ppm}$, and  was measured at $\langle Q^2\rangle=0.02357 \pm 0.00010$ GeV$^2$, $\langle \theta_{\rm lab}\rangle =7.61^\circ \pm 0.02^\circ$, 
and $\langle E_{\rm lab}\rangle=1.157 $ GeV with the Q$_{\rm weak}$ apparatus at Jefferson Lab. Predictions 
using a simple 
Born approximation as well as more sophisticated distorted-wave calculations are in good agreement with this result.
From this asymmetry  the $^{27}$Al neutron radius $R_{n} = 2.89\pm0.12 \text{ fm}$ was determined using a many-models correlation technique. The corresponding neutron skin thickness $R_{n}-R_{p} = -0.04\pm0.12 \text{ fm}$ is small, as expected for a light nucleus with a neutron excess of only 1. This result thus serves as a successful benchmark for electroweak determinations of neutron radii on heavier nuclei.  
A tree-level approach was used to extract
the $^{27}$Al weak radius $R_{\rm w} = 3.00\pm0.15 \text{ fm}$, and the weak skin thickness $R_{\rm wk}-R_{\rm ch} = -0.04\pm0.15 \text{ fm}$.
The weak form factor at this $Q^2$ is $F_{\rm wk}=0.39 \pm 0.04$.
\end{abstract}

\maketitle

As beam properties and experimental techniques have  improved over the last two decades, so has the precision of parity-violating (PV)  asymmetry measurements in elastic electron %elastic
scattering. 
These experiments initially focused on carbon~\cite{Bates12C}, then hydrogen   
and helium  
targets to study strange quark form factors
\cite{Armstrong:2012bi}. The improving precision of these experiments has led to standard model tests ~\cite{PhysRevLett.111.141803,Qweak.Nature}, and even more recently neutron radius determinations in heavy nuclei ~\cite{PREX:2021umo,Horowitz:2013wha} %up to $^{208}$Pb 
which impact our understanding of the structure and composition of
neutron stars~\cite{PhysRevLett.126.172503}. 

%One such recent experiment determined 
The proton's weak charge was determined 
in the Q$_{\text{weak}}$ experiment  ~\cite{Qweak.Nature,Carlini2019}
by measuring the PV asymmetry in $\vec{e}p$ elastic scattering with high precision at low $Q^2$. By far the largest background in that experiment ($\approx 24\%$) %(f_i)(A_i)/(1-ftot) was 37 ppb in run 2 out of -226.5 ppb total
came from the aluminum alloy cell that contained the hydrogen. To accurately account for that background, precise
additional asymmetry measurements were made on aluminum interspersed between data taking on hydrogen.

Those same aluminum asymmetry results that served to account for background in the Q$_{\text{weak}}$ experiment  have been further analyzed
in this work to isolate the $^{27}$Al asymmetry $A_{\rm PV}$  for elastic electron scattering  at $Q^2= 0.02357$ GeV$^2$. A successful comparison with theory~\cite{PhysRevC.89.045503} would provide additional confidence in  the empirical background subtraction 
used in the Q$_{\text{weak}}$ experiment~\cite{Qweak.Nature}. 

However, the most important aspect of the first $^{27}$Al $A_{\rm PV}$ measurement presented here is the test case it provides for the electroweak (EW) technique~\cite{PhysRevC.63.025501} used to determine the neutron radius $R_n$ of a complex nucleus in $\vec{e}A$ scattering. In conjunction with the more easily-determined proton radius $R_p$, this also delivers the neutron skin $R_n-R_p$.  

For a light complex nucleus like $^{27}$Al with a neutron excess of only 1, we expect the neutron skin to be very thin. If this naïve expectation is confirmed by our measurement, it would serve as a benchmark for the application of the EW technique to heavier nuclei like $^{208}$Pb, where the resulting neutron skin can be related to neutron star physics~\cite{PhysRevLett.126.172503}. The EW technique has recently been applied to $^{208}$Pb~\cite{PREX:2021umo} and the resulting neutron skin 
%thickness 
was found to be in some tension with earlier non-EW results ~\cite{EW_nonEW,Steiner} which favor a thinner skin.
%, 
%as well as with recent astrophysical analyses from %LIGO~\cite{PhysRevLett.121.161101} and %NICER~\cite{PhysRevLett.126.172503} which are also more %compatible with the non-EW results. 
The benchmark of the EW technique which our result can provide is especially important in light of this observed tension. 

     Beyond providing the $^{27}$Al asymmetry $A_{\rm PV}$, neutron radius $R_n$, and neutron skin 
     thickness 
     $R_n - R_p$, we also report the $^{27}$Al weak form factor $F_{\rm wk}$ at our $Q^2$, the $^{27}$Al weak radius $R_{\rm wk}$ and weak skin thickness $R_{\rm wk} - R_{\rm ch}$, where $R_{\rm ch}$ is the charge radius. $R_{\rm wk}$ should closely track the neutron radius because the weak charge comes primarily from the neutrons – the proton’s weak charge is much smaller~\cite{Qweak.Nature}.

A PV asymmetry is a non-zero difference between 
differential cross sections $\sigma_{\pm}(\theta)$ measured with a beam polarized parallel $(+)$ or anti-parallel $(-)$ to its incident momentum. In the Born approximation the elastic $\vec{e}-^{27}$Al asymmetry can be expressed \cite{PhysRevC.89.045503} as
\begin{equation}\label{eq:pv_asymmetry}
    A_{\text{PV}} = \frac{\sigma_{+}(\theta)-\sigma_{-}(\theta)}{\sigma_{+}(\theta)+\sigma_{-}(\theta)} \approx \frac{-G_{F}Q^{2}Q_W}{4\pi\alpha Z \sqrt{2}}\frac{F_{\text{W}}(Q^{2})}{F_{\text{EM}}(Q^{2})},
\end{equation}
\noindent where $G_{F}$ is the Fermi constant, $\alpha$ is the fine structure constant, $-Q^2$ is the four-momentum transfer squared, $Q_W=-12.92\pm 0.01$  
is the predicted~\cite{Zyla:2020zbs} weak charge of $^{27}$Al including all radiative corrections, and $Z$ is the atomic number of $^{27}$Al. $F_{\text{W}}(Q^{2})$  and $F_{\text{EM}}(Q^{2})$ are the weak and electromagnetic (EM) form factors for $^{27}$Al,  normalized to unity at $Q^2=0$.

This measurement was conducted in Hall C of Jefferson Lab 
using the Q$_{\text{weak}}$ experimental apparatus~\cite{ALLISON2015105} and the polarized electron beam of the CEBAF accelerator. 
The helicity of the polarized electron beam was selected at a rate of 960 Hz, allowing the beam to be produced in a sequence of “helicity quartets”, either $(+--+)$ or $(-++-)$, with the pattern chosen pseudorandomly at 240 Hz.
In addition, 
every 
$8$ hours an insertable half-wave plate (IHWP) was placed in or out of the source laser's path to reverse the polarization direction. A `double Wien' spin rotator was also used to reverse the electron spin direction twice during the $^{27}$Al data-taking.

A $60 \text{ $\mu$A}$ longitudinally polarized $1.16 \text{ GeV}$  electron beam  was incident on a $3.68 \text{ mm}$ thick by $2.54 \text{ cm}$ square 7075-T651 aluminum alloy target. This target was machined from the same lot of material used for the LH$_2$ target window components of the weak charge measurement,
so it could also be used to account for the
background  aluminum asymmetry that contaminated the measured hydrogen asymmetry~\cite{PhysRevLett.111.141803,Qweak.Nature}. Other elements in this alloy, as determined during a post-experiment assay, include: Zn ($5.87 \text{ wt\%}$), Mg ($2.63 \text{ wt\%}$), Cu ($1.81 \text{ wt\%}$), 
and other ($0.47 \text{ wt\%}$).
%Cr ($0.19 \text{ wt\%}$), Fe ($0.11 \text{ wt\%}$), Si ($0.09 \text{ wt\%}$), Mn ($0.05 \text{ wt\%}$), and Ti ($0.03 \text{ wt\%}$).

\begin{figure}[bbth]
    \centering
    \includegraphics{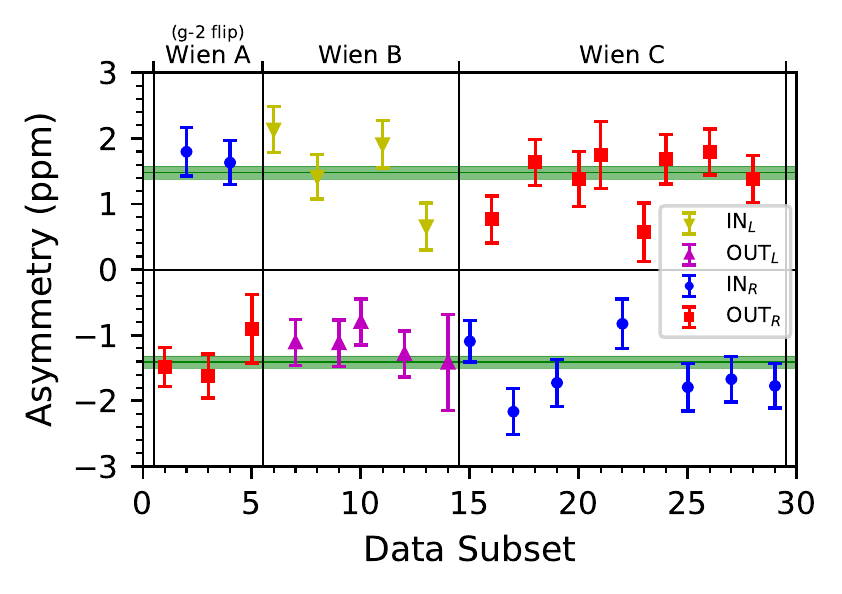}
    \caption{Raw %unregressed 
    asymmetries (statistical errors only)  plotted against 8-hour IHWP IN or OUT  ``data subsets"  (lower axis), and monthly $L$ or $R$ Wien spin %filter 
    rotator orientation (upper axis). 
    The configuration consistent with Eq.~\ref{eq:pv_asymmetry}
    is given by Wien Left and IHWP IN, i.e.\ IN$_L$, which is equivalent to OUT$_R$. The opposite sign asymmetry arises when either the Wien or the IHWP is flipped, but not both. During Wien A, there was an additional ($g-2$) spin flip which arose from running the JLab recirculating linac with 2 passes at half the gradient instead of 1 pass with full gradient. The green lines (bands) denote weighted averages (uncertainties) of the positive and negative asymmetries. 
}
    \label{fig:raw_asymmetry_data}
\end{figure}

Electrons scattered from the target were first 
selected by a series of three collimators and were then focused by a toroidal magnetic field onto an azimuthally-symmetric array of eight synthetic-quartz Cherenkov detectors, each with a 2-cm-thick lead preradiator. The  polar-angle ($\theta$) acceptance was $5.8^{\circ}$ to $11.6^{\circ}$, the azimuthal-angle acceptance was  $49\text{\%}$ of $2\pi$, and the energy acceptance was large:  $\approx 150$ MeV. Cherenkov light generated in the quartz from the passing electrons was collected by photomultiplier tubes (PMTs) attached to each end of each detector in the array. The current from the PMTs was integrated over each helicity state, normalized to the beam current and then averaged together to form the raw asymmetry $A_{\text{raw}}$, as shown in Fig.~\ref{fig:raw_asymmetry_data} and %given in
Tab.~\ref{tab:raw_asymmetry_stats}.
%
%Word Length: 123
\begin{table}[tbh]
    \centering
    \caption{%Values of the 
    Time-averaged raw %(unregressed) 
    asymmetries and their statistical uncertainties. $A_{\text{raw}}$  is the weighted average of the sign-corrected raw asymmetries NEG and POS. NULL is the arithmetic average of NEG and POS.  
    The $\chi^{2}$ per degree of freedom and associated probabilities are given for each type of average.
    }
    \label{tab:raw_asymmetry_stats}
    \begin{tabular}{cccc}
        \toprule
        {Average} & {Asymmetry(ppm)} & {$\chi^{2}/d.o.f.$} & {$\chi^{2}$ Prob.}\\ \midrule
        NEG & -1.407 $\pm$ 0.093 & 1.26 & 0.225 \\
        POS & 1.480 $\pm$ 0.099 & 1.62 & 0.073 \\
        NULL & 0.036 $\pm$ 0.068 & -- & -- \\
        $A_{\text{raw}}$ & 1.441 $\pm$ 0.068 & 1.39 & 0.082 \\ \bottomrule
    \end{tabular}
\end{table}
Several small 
systematic corrections were applied to $A_{\text{raw}}$ to derive a measured asymmetry $A_{\text{msr}}$:
\begin{equation}\label{eq:measured_asymmetry}
    A_{\text{msr}} = A_{\text{raw}} + A_{\text{BCM}} + A_{\text{reg}} + A_{\text{BB}} + A_{\text{L}} + A_{\text{T}} + A_{\text{bias}},
\end{equation}
\noindent 
where $A_{\text{BCM}}$ is a beam-current monitor (BCM) normalization uncertainty, $A_{\text{reg}}$ is a helicity-correlated beam-motion correction, $A_{\text{BB}}$ is a beam-line background correction, $A_{\text{L}}$ is a non-linearity correction, $A_{\text{T}}$ is a residual transverse-asymmetry correction, and $A_{\text{bias}}$ is a rescattering bias correction. Each of these corrections is discussed below.

The raw asymmetry charge normalization adopted the same technique and BCMs as used in the weak charge measurement~\cite{Qweak.Nature}, 
leading to a correction
of $A_{\text{BCM}}=0.0\pm2.1 \text{ ppb}$, dominated
by the BCM accuracy.%, which was adopted from calculations performed in the weak charge analysis~\cite{Qweak.Nature}.

Helicity-correlated variations in the beam position and energy also required a correction $A_{\text{reg}}=0.4\pm1.4 \text{ ppb}$. This was determined with a linear regression method~\cite{PhysRevLett.111.141803,Dissertation:KurtisBartlett}, 
to correct the effects of natural beam-motion using helicity-correlated differences measured with different beam-position monitors. 

Electrons in the 
beam halo interacted with beam-line components causing a false asymmetry. Auxiliary detectors 
placed close to the beam line 
were used to form a
correlation with the main detectors 
to correct for this false asymmetry. 
The overall correction was $A_{ \text{BB}}=-4.7\pm6.6 \text{ ppb}$.

Non-linearity effects in the main detector PMTs and BCMs used for asymmetry normalization were quantified in bench-top tests. The correction for this effect was 
$A_{\rm L}=-2.0\pm7.0 \text{ ppb}$~\cite{Dissertation:KurtisBartlett,Dissertation:WadeDuvall}.

Any residual transverse components to the beam polarization will cause a parity-conserving azimuthal variation in the asymmetry, which coupled with imperfections in the azimuthal symmetry of the detectors may lead to a false asymmetry. This was measured using transversely polarized beam  \cite{QWeak:2021jew} and scaled to the measured azimuthal variation in the present data, leading to a correction
$A_{\rm T}=-3.4\pm8.8 \text{ ppb}$ \cite{Dissertation:KurtisBartlett}. 

As described in earlier publications~\cite{Qweak.Nature, Carlini2019}, lead pre-radiators placed in front of the main detectors were needed to reduce low-energy backgrounds. However, scattered electrons 
with spins precessing from longitudinal to transverse in
the spectrometer 
magnetic field
acquired an analyzing power from  Mott scattering in the lead, which led to a correction of $A_{\rm bias}=4.3\pm3.0 \text{ ppb}$.

Determination of a purely elastic asymmetry $A_{\text{PV}}$ required additional corrections for beam polarization, background asymmetries, and a combination of radiative and acceptance corrections: 
\begin{equation}\label{eq:pv_asymmetry_corrections}
    A_{\text{PV}} = R_{\text{tot}}\frac{A_{\text{msr}}/P - \sum_{i}f_{i}A_{i}}{1-\sum_{i}f_{i}} ,
\end{equation}
\noindent where $R_{\text{tot}} = 
0.9855 \pm 0.0087$, determined primarily by simulation~\cite{Qweak.Nature}, accounts for the radiative and finite acceptance effects,
$f_{i}$ is the signal fraction of a particular background asymmetry, and $A_{i}$ is its corresponding asymmetry. These can be found in Tab.~\ref{Tab:Corrections}.

The beam polarization was monitored
continuously
using a Compton polarimeter~\cite{Compton} and periodically with dedicated measurements using a M\o{}ller polarimeter~\cite{Moller}. Both were found to 
agree~\cite{MAGEE2017339} during the experiment and yielded a combined polarization  of $P=88.80\pm0.55 \text{\%}$.

Non-elastically scattered electrons entering the large acceptance of the apparatus contaminated the measured asymmetry with backgrounds which had to be estimated and subtracted in Eq.~\ref{eq:pv_asymmetry_corrections}. Non-elastic processes considered in this analysis include quasi-elastic, single-particle and collective  excitations, and inelastic scattering with a $\Delta$ in the final state. Correction for each of these backgrounds required knowledge of 
the fraction of events that fell into the acceptance, $f_i$, derived from the cross section of each process at the kinematics of the experiment, and $A_i$, the asymmetry for each process. Both of these were determined using models and/or experimental data from previous measurements.
The relevant dilutions for each of these background processes were reported in~\cite{QWeak:2021jew}.

The quasi-elastic asymmetry $A_{\rm QE}$ was estimated for $^{27}$Al from a relativistic Fermi gas model \cite{Horowitz:1993nb}, with a conservative 50\% relative uncertainty. 

The inelastic asymmetry $A_{\rm inel}$ was determined by 
dropping the spectrometer magnetic field to about 75\% of its nominal value to move the inelastic events onto the detectors. The corresponding polarization-corrected $^{27}$Al asymmetry
\begin{equation} \label{eq:Ainel_scaling}
A^{75}=f^{75}_{\rm el} A^{75}_{\rm el} + f^{75}_{\rm inel} A^{75}_{\rm inel}=1.36 \pm 0.97 \; {\rm ppm}
\end{equation} 
was briefly measured, with $f^{\rm 75}_{\rm inel}$ 
estimated from simulation to be ($20\pm 5)\%$ on top of the elastic tail,
and $A^{75}_{\rm el}$ scaled down from its value at full field by 1.181, the ratio of the corresponding $Q^2$ at each field. A value for $A_{\rm inel}=-0.58\pm5.83$ ppm at full field was obtained by solving Eq.~\ref{eq:Ainel_scaling} for $A^{75}_{\rm inel}$ and then scaling up by the $Q^2$ ratio.

Following the work of \cite{PhysRevC.89.045503}, the asymmetry for the giant dipole resonance was estimated using the Born approximation for an $N=Z$ nucleus,  with a negative sign $A_{\rm GDR}=-2.2\pm1.1$ ppm appropriate for this isovector transition, and a conservative 50\% relative uncertainty.

Asymmetries $A_{\rm nucl} \approx 2.5 $ ppm for the 11 strongest excited states of $^{27}$Al up to 7.477 MeV were also obtained using the Born approximation for elastic scattering,  with small corrections made for the acceptance-averaged $Q^2$.
%we made all the excited state asymmetry uncertainties with W.u.'s > 7.8 or excited by T=0 probes 50%, and all the others 100%.
%All excited states  with large E2 transition rates or excited by T=0 probes were assigned uncertainties of 50\%, and all the others 100\%.
States with large E2 transition rates or which were strongly populated by  $T=0$ probes %$(\alpha,\alpha^\prime)$ 
were assumed to be isoscalar and assigned  50\%  uncertainties. 
 The remaining states were assumed to be isovector. Since the sign of the asymmetry depends on whether those isovector states were proton or neutron excitations, a 200\%  uncertainty was used to encompass both possibilities.

For the asymmetries $A_{\rm alloy}$ associated with the contaminant elements in the alloy used for the target, the Born approximation calculation was again used as described in \cite{PhysRevC.89.045503} for each of the dominant six elements. These calculations include Coulomb distortions, but assume spherically symmetric proton and neutron distributions, so only include the leading multipole term. As before, 50\% uncertainties were used.

Background contributions from pions, neutrals, and the beamline were negligible, and are discussed in \cite{Dissertation:KurtisBartlett}.

\begin{table}[!htb]
 \caption{Corrections applied to obtain the final asymmetry $A_{\rm PV}$ and their corresponding contributions to the systematic uncertainty. The total systematic uncertainty is the quadrature sum of these uncorrelated uncertainties. %{\color{red} Updated 10/6/2021.}  
 }
 \label{Tab:Corrections}
    \centering
    \begin{tabular}{lcc }
\hline
   Quantity & Value   & $\Delta A_{\rm PV}/A_{\rm PV}$ (\%)  \\
%      &     (ppm)  & (\%) \\
   \hline 
% & &  \\

$A_{\rm msr}$: & $ 1.436 \pm 0.014$ ppm & 1.0\\
$P$:  &   $0.8880 \pm 0.0055$  & 0.7 \\
$R_{\rm tot}$: & $ 0.9855 \pm 0.0087$ & 0.9\\
%   $R_{\rm av}$: Acceptance averaging &  $0.9862 \pm 0.0036$    & 0.4\\
% $R_l$: Electronic non-linearity & $1.0014 \pm 0.0050$  & 0.6 \\   
% $A_{\rm fit}$: Fitting  & $0 \pm 0.042$ ppm  & 0.6\\
% $A_{\rm reg}$: Linear Regression & $0 \pm 0.002$ ppm  & 0.3\\
% $A_{\rm bias}$: Rescattering Bias & $0.125 \pm 0.041$ ppm  & 0.6 \\
%
$f_{\rm QE}$: & $21.2 \pm 2.9$ \%  &  5.0 \\
$A_{\rm QE}$:& $-0.34 \pm 0.17$ ppm  & 2.4 \\
$f_{\rm nucl}$: & $3.83 \pm 0.23$ \%  & 0.1\\
$A_{\rm nucl}$: & $2.58 \pm  1.40$ ppm   &   3.6\\
$f_{\rm inel}$: & $0.665\pm 0.099$ \% & 0.2 \\
$A_{\rm inel}$:& $-0.58 \pm 5.83$ ppm  & 2.6\\
$f_{\rm alloy}$:  & $5.41 \pm 0.34$ \% &   0.1 \\
$A_{\rm alloy}$:  & $1.90 \pm 0.58$ ppm  &  2.1\\%
 $f_{\rm pions}$: & $0.06\pm 0.06$ \%  & 0.1 \\
 $A_{\rm pions}$:& $0\pm 20$ ppm  & 0.8 \\
   $f_{\rm neutral}$: & $0\pm 0.45$ \%  & 0.1 \\
   $A_{\rm neutral}$:& $1.7\pm 0.2$ ppm  & 0.0 \\
 $f_{\rm beamline}$: & $0.69\pm 0.06$ \%  & 0.1 \\
 %$A_{\rm beamline}$:& $0\pm 0$ ppm  & 0.0 \\
 % DSA: there is no A_beamline correction, as that is done in A_BB false asymmetry earlier. 
$f_{\rm GDR}$:  & $0.045 \pm 0.023$ \% &   0.1 \\
$A_{\rm GDR}$:  & $-2.22 \pm 1.11$ ppm  & 0.0\\

        \hline
Total Systematic &  &   7.6  \\
\end{tabular}
\end{table}

\begin{figure}[t]
    \centering
    \includegraphics{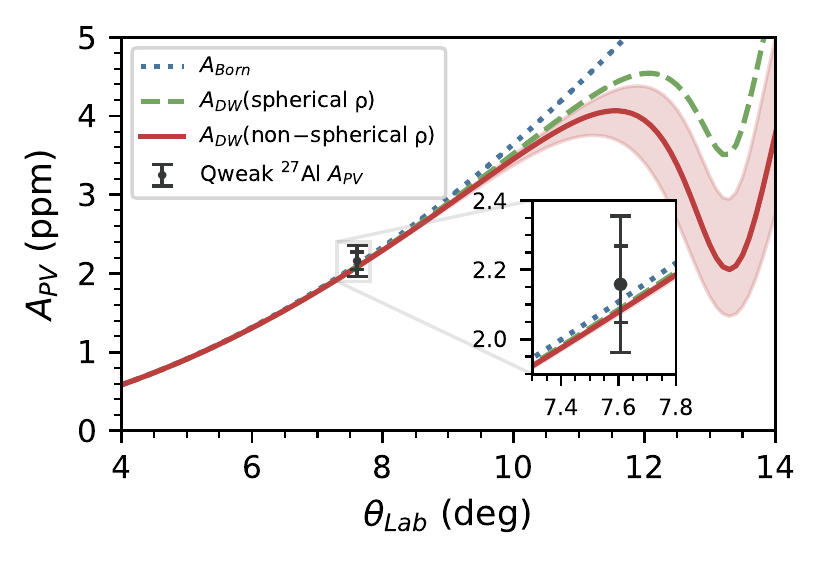}
    \caption{Parity-violating asymmetry vs. laboratory scattering angle. The measured value is shown with statistical (inner error bar) and total (outer error bar) uncertainties. The theoretical prediction \cite{ PhysRevC.89.045503} at our beam energy is shown for spherically symmetric neutron and proton densities in Born approximation (blue dots), for a distorted wave calculation with
spherical densities (dashed green line) and the full calculation with
non-spherical proton density (red solid line). The red shaded band indicates
nuclear structure and Coulomb distortion uncertainties.
}
\label{fig:asymmetry_measurement_theory_compare}
\end{figure}

After all corrections, the elastic $^{27}$Al asymmetry is
%Word Length: 16
\begin{equation} \label{eq:APV_Result}
%    A_{\text{PV}} = 1.927 \pm 0.091\text{(stat)} \pm 0.148\text{(syst)} \text{ ppm}
    A_{\text{PV}} = 2.16 \pm 0.11\text{(stat)} \pm 0.16\text{(syst)} \text{ ppm}
\end{equation}
\noindent at $Q^{2} = 0.02357\pm0.00010 \text{ GeV}^{2}$, which corresponds to $\langle \theta_{\text{Lab}} \rangle = 7.61^{\circ} \pm 0.02^{\circ}$. This result, the first on $^{27}$Al, agrees well %within its 1$\sigma$ uncertainty 
with previously published distorted wave Born calculations~\cite{PhysRevC.89.045503} as shown in Fig.~\ref{fig:asymmetry_measurement_theory_compare}. 

\begin{figure}[t]
    \centering
    \includegraphics{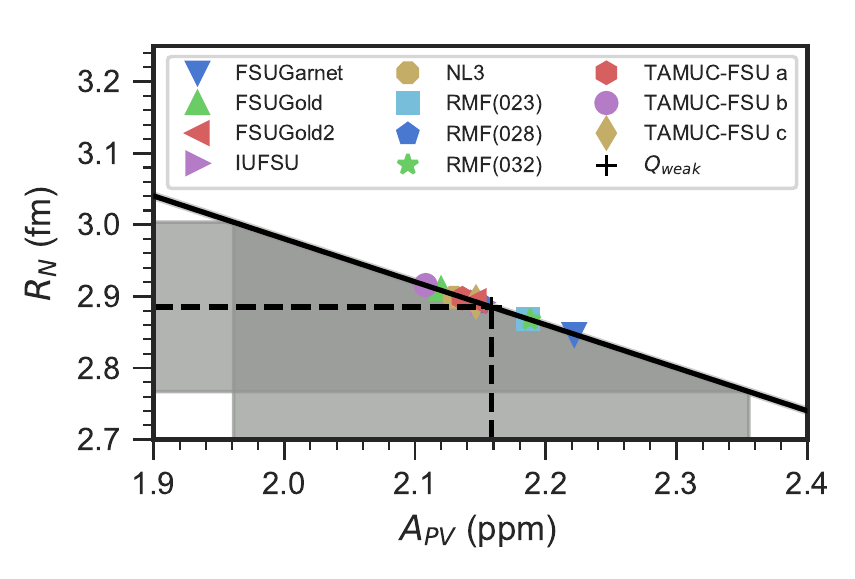}
    \vspace*{-0.75cm}
    \caption{%{\color{red} Needs work.} 
    %Some of the many m
    Models (symbols indicated in the legend) used to establish the correlation (Eq.~\ref{eq:neutron_asymmetry_model_relation}, and solid black line) between the $^{27}$Al $A_{\rm PV}$ and its neutron radius $R_n$. 
    The dashed black lines indicate where on the many-models correlation plot the central value of our asymmetry determines $R_n$. The shaded bands indicate the %statistical and 
    total uncertainty associated with our result.
    }
    \label{fig:neutron_radius_extraction}
\end{figure}

The neutron distribution radius $R_{n}$ was determined using a many-models correlation method first employed by the PREX collaboration~\cite{PhysRevLett.108.112502}. A selection of relativistic mean-field models~\cite{Article:Modelfsugold,Article:Modelfsugold2,Article:Modelnl3,Article:Modelnl3supp,Article:Modeliufsu,Article:Modeltamu,Article:Modelfsugarnet} were chosen based on their
ability to reasonably predict 
several nuclear structure observables: nucleon binding energies, charge radii, and strengths of isoscalar and isovector giant resonances in selected nuclei. 
The relationship between $R_n$ and $A_{PV}$ was found to be 
\begin{equation}\label{eq:neutron_asymmetry_model_relation}
    R_{n} = (-0.6007 \pm 0.0002)\frac{A_{\text{PV}}}{\text{ppm}} + (4.1817 \pm 0.0011) \;{\rm fm}
\end{equation}
%Rn = (−0.6007±0.0002) APV/ppm + (4.1817±0.0011) fm, (6)
with correlation coefficient 0.997.  Using this relation our final asymmetry yielded $R_{n} = 2.89\pm0.12 \text{ fm}$, see Fig.~\ref{fig:neutron_radius_extraction}.

To determine the neutron skin %thickness
$R_{n}-R_{p}$, we use the proton distribution radius $R_{p}$ following Ref.~\cite{Ong:2010gf} for spherical nuclei,
\begin{equation} \label{eq:Rp}
\begin{split} 
    R_p &= \left(R^2_{\rm ch}-\langle r_p^2\rangle  - \frac{N}{Z}\langle r^2_n\rangle - \frac{3}{4m_N^2}-\langle r^2_{\rm so}\rangle \right)^{1/2}\\
    & =2.925\pm0.007\;\text{fm} ,
% & =2.962\pm0.038\;\text{fm} , %new Rch&rp
    \end{split}
\end{equation}
where $m_N$ is the nucleon mass, and $N$ denotes the number of neutrons. Here and below we use an $^{27}$Al charge radius $R_{\rm ch}=3.035\pm0.002$ fm~\cite{DEVRIES1987495}, and correct for the proton charge radius $\langle r_p\rangle =0.8751\pm0.0061$ fm~\cite{PDG2017}, the neutron charge radius $\langle r^2_n\rangle=-0.1161\pm 0.0022$ fm$^2$ ~\cite{Zyla:2020zbs},  and a spin-orbit nuclear charge correction $\langle r^2_{\rm so}\rangle =-0.017$~fm$^2$  following 
%the prescription in
\cite{Ong:2010gf}.  For consistency these parameters must be the same as those used to extract $R_n$ using Eq.~\ref{eq:neutron_asymmetry_model_relation}. 
The neutron skin %thickness
is $R_{n}-R_{p} = -0.04 \pm 0.12 \text{ fm}$, confirming %at the $0.4 \sigma$ level 
the naive expectation for a light nucleus such as $^{27}$Al where $N \approx Z$ that the neutron skin should be close to zero within our uncertainty. To illustrate the sensitivity of $R_p$ to its input parameters, using other recent values for $\langle r_p\rangle$~\cite{Zyla:2020zbs} and $R_{\rm ch}$~\cite{27Al_Rch} would only raise $R_p$ by 1\%, which is small compared to our 4.2\% precision for $R_n$.

In order to proceed to estimates of  %approximate results for
electroweak (EW) observables to which this experiment is sensitive (see Tab. ~\ref{tab:EW_Observables}), we follow the Born approximation (tree-level) formulation presented in~\cite{Koshchii:2020qkr}. Although this leads only to approximate EW  results, the 9.1\% precision of our asymmetry is large enough to blunt the need for a more precise treatment. In addition, Fig.~\ref{fig:asymmetry_measurement_theory_compare} shows that the Born approximation accurately predicts our asymmetry. Moreover, the relatively low $Z$ of $^{27}$Al reduces the corrections from Coulomb distortions ($\propto Z$) relative to a heavier nucleus like Pb.

Following Ref.~\cite{Koshchii:2020qkr}, we introduce a term  $\Delta$ which accounts for hadronic and nuclear structure effects at $Q^2 >0$:
\begin{equation} \label{eq:Delta}
    \Delta \equiv \frac{F_{\rm wk}(Q^2)}{F_{\rm EM}(Q^2)}-1 = \frac{A_{\rm PV}}{A_0} \frac{Z}{Q_W}-1,
\end{equation}
where  $ A_0=-G_F Q^2/(4\pi\alpha\sqrt{2})$. 
Inserting our $A_{\rm PV}$ result (Eq.~\ref{eq:APV_Result}) into either Eq.~\ref{eq:pv_asymmetry} or 
Eq.~\ref{eq:Delta}, and using an $F_{\rm EM}=0.384 \pm 0.012$ calculated following the prescription outlined in \cite{STOVALL1967513}, we obtain a weak form factor $F_{\rm wk}(Q^2=0.0236\text{ GeV}^2) = 0.393\pm 0.038$. 
The $F_{\rm EM}$  calculation (corrected for small Coulomb distortions) is good to about 3\%~\cite{STOVALL1967513}, which we verified by comparing with differential cross section data ~\cite{PhysRevC.9.1861}. 

With our $A_{\rm PV}$ result, 
$\Delta = 0.025\pm0.094$. To lowest order in $Q^2$, $R_{\rm wskin}\equiv R_{\rm wk}-R_{\rm ch}=-3\Delta /(Q^2 R_{\rm ch})$~\cite{Koshchii:2020qkr},
from which we obtain $R_{\rm wskin}=-0.04\pm0.15$ fm, consistent as expected with our small neutron skin result. Employing the $R_{\rm ch}$ introduced earlier, $R_{\rm wk}=3.00\pm0.15$~fm. The relative difference between the weak and charge radii $\lambda\equiv(R_{\rm wk}-R_{\rm ch})/R_{\rm ch}=-1.3\%\pm 5.0\%$.

\begin{table}[hb]
    \centering
    \caption{%Values of the 
    Derived $^{27}$Al Observables
    }
    \label{tab:EW_Observables}
    \begin{tabular}{cccc}
        \toprule
        {Observable} & {Value } & Uncertainty & Units\\ \midrule
        $R_n$ & 2.89  & 0.12  & fm\\
        %$R_p$ & 2.925  & 0.018 & fm \\
        $R_n-R_p$ & -0.04  & 0.12 & fm \\
        %$F_{EM}(Q^2=0.0236$ GeV$^2$) & 0.384  & 0.012 &  \\
        $F_{\rm wk}(Q^2=0.0236$ GeV$^2$) & 0.393  & 0.038 &  \\
        $\Delta=Z A_{\rm PV}/(A_0 Q_W) -1$ & 0.025  & 0.094 &  \\
        $R_{\rm wskin}=-3\Delta/(Q^2 R_{\rm ch})$  & -0.04 & 0.15 & fm \\
        $R_{\rm wk}=R_{\rm wskin}+R_{\rm ch}$ & 3.00  & 0.15 & fm \\
        $\lambda\equiv (R_{\rm wk}-R_{\rm ch})/R_{\rm ch}$ & -1.3  & 5.0 & \% \\
         \bottomrule
    \end{tabular}
\end{table}

In conclusion, the agreement between predictions~\cite{PhysRevC.89.045503} and this first measurement of 
the  elastic asymmetry on $^{27}$Al  supports the background procedures used in the Q$_{\rm weak}$ experiment~\cite{Qweak.Nature} on hydrogen. %were valid.}
The  tree-level EW results obtained above for $R_{\rm wk}$ and $R_{\rm wskin}$ 
are consistent with broad expectations for a low-$Z$ nucleus with $N \approx Z$ such as $^{27}$Al. 
Similarly, our $^{27}$Al neutron skin is %also
close to zero, 
as expected, 
%matching naive expectations for a light $N \approx Z$ nucleus, 
providing some validation and a benchmark for the application of the many-models approach and EW technique ~\cite{PhysRevC.63.025501} to the measurement of heavier nuclei~\cite{PREX:2021umo,Horowitz:2013wha,PhysRevLett.108.112502}. 

This is especially interesting in light of the tension which exists~\cite{EW_nonEW,PhysRevLett.127.232501,PhysRevC.104.065804,Most:2021ktk} between the recent EW neutron skin determination $R_n-R_p=0.283 \pm 0.071$~fm for $^{208}$Pb~\cite{PREX:2021umo}, and the 2012 average of several disparate but self-consistent non-EW determinations $R_n-R_p=0.184 \pm 0.027$ fm~\cite{Steiner}.
The older non-EW determinations have come under additional scrutiny and even some criticism recently~\cite{Thiel_2019}. However, we note that they appear to be more consistent with the latest constraints on neutron star properties from LIGO and Virgo (especially for the tidal deformability) 
~\cite{PhysRevLett.121.161101}, from NICER~\cite{PhysRevLett.126.172503}, and  astrophysical models in general.

%%%%%%%%%%%%%%%%%%%%%%%%%%%%%%%%%%%%%%%%%%%
\begin{acknowledgements}

We thank the staff of Jefferson Lab, in particular the accelerator operations staff, the radiation control staff, as well as the Hall C technical staff for their help and support. We are also grateful for the contributions of our undergraduate students. We thank TRIUMF for its contributions to the development of the spectrometer and integrated electronics, and BATES for its contributions to the spectrometer and Compton polarimeter. 
We also thank T.W.~Donnelly for helpful discussions.
This material is based upon work supported by the U.S. Department of Energy (DOE), Office of Science, Office of Nuclear Physics under contract DE-AC05-06OR23177.
Construction and operating funding for the experiment was provided through the DOE, the Natural Sciences and Engineering Research Council of Canada (NSERC), the 
Canada Foundation for Innovation (CFI), and the National Science Foundation (NSF) with university matching contributions from William \& Mary, Virginia Tech, George Washington University and Louisiana Tech University. 
\end{acknowledgements}

\bibliographystyle{apsrev}
\bibliography{references.bib}

\end{document}